\documentclass[aps,prb,twocolumn,groupedaddress,showpacs]{revtex4}

\usepackage{graphicx}
\usepackage{amsmath}
\usepackage{amssymb}
\usepackage{bm} 

\newcommand{\abs}[1]{\left\lvert#1\right\rvert}
\newcommand{\av}[1]{\left\langle#1\right\rangle}

\newcommand\etal{\emph{et al.\ }}

\DeclareMathOperator\Tr{tr}

\newcommand\figwidth{3.375in}

\providecommand{\openone}{\leavevmode\hbox{\small1\kern-3.8pt\normalsize1}}

\newcommand\NL{\ensuremath{N}}
\newcommand\NR{\ensuremath{M}}
\newcommand\Ntot{\ensuremath{K}}

\newcommand\win[1][]{\ensuremath{w^{#1}}}
\newcommand\wout{\ensuremath{w^{\text{out}}}}
\newcommand\tildewout{\ensuremath{\tilde{w}^\text{out}}}
\newcommand\CUE{\text{CUE}}
\newcommand\CSE{\text{CSE}}

\providecommand\Sdag{S^\dagger}

\providecommand\Ta{t}

\providecommand\jr[1][]{j_w^#1}
\providecommand\Jr[1]{j_\text{in}^#1}
\providecommand\vecjr{\vec{j}_w}
\providecommand\g{j_\text{out}}
\providecommand\p{p_w}

\providecommand\vecp{\vec{p}_w}

\graphicspath{{Figs/}}

\begin{document}
\title{Fluctuations of spin transport through chaotic quantum dots with spin-orbit coupling}
\author{Jacob J.\ Krich}
\affiliation{Physics Department, Harvard University, Cambridge, MA 02138}
\date{\today}

\begin{abstract}
    As devices to control spin currents using the spin-orbit interaction are proposed and implemented, it is important to understand the fluctuations that spin-orbit coupling can impose on transmission through a quantum dot. Using random matrix theory, we estimate the typical scale of transmitted charge and spin currents when a spin current is injected into a chaotic quantum dot with strong spin-orbit coupling. 
    These results have implications for the functioning of the spin transistor proposed by Schliemann, Egues, and Loss. We use a density matrix formalism appropriate for treating arbitrary input currents and indicate its connections to the widely used spin-conductance picture. We further consider the case of currents entangled between two leads, finding larger fluctuations.
\end{abstract}
\maketitle

\section{Introduction}

There has been much recent progress in the creation and control of spin currents. There have been demonstrations and proposals for producing spin-polarized currents both with \cite{Kikkawa99,Stich07,Eto05,Bardarson07,Krich08} and without time reversal symmetry (TRS).\cite{Lou07,Frolov09}  Recent progress in measuring and controlling the spin-orbit coupling in semiconductor heterostructures \cite{Koralek09,Lechner09,Studer09} promises to enable a range of spintronic applications relying on the spin-orbit interaction. As such devices are considered and developed, it is important to understand the role of coherent mesoscopic fluctuations in these systems. In this paper, we consider the effects of injecting either a spin-polarized current or a pure spin current into a two-dimensional ballistic region with strong spin-orbit coupling and consider the scale of the fluctuations of charge and spin currents transmitted through such a device.

For example, these effects could be important for the Schliemann-Egues-Loss spin field effect transistor (SFET) proposal.\cite{Schliemann03} In such a SFET, spin-polarized electrons are injected into a region (e.g., a diffusive wire or a quantum dot) with spin-orbit coupling. In the ``on'' state of the device, the Rashba \cite{Bychkov84} and $k$-linear Dresselhaus \cite{Dresselhaus55,Winkler03} spin-orbit couplings are tuned to be equal, and the spin polarization does not decay as the electrons cross the region, but instead undergoes a controlled rotation.\cite{Schliemann03} In the ``off'' state, the Rashba and $k$-linear Dresselhaus strengths are tuned to be different, and the spin polarization is lost while traversing the region due to the random spin rotations experienced by electrons traversing different trajectories through the dot. Ideally, the on state has a fully spin-polarized current exiting the device and the off state has no spin polarization in the exit current. For coherent 2D quantum systems, however, the decay of the spin current in the off state relies on having a sufficiently large number of channels to average together. In the 1D limit, with two ideal one-dimensional wires, each having only one propagating mode, a fully spin-polarized current injected into the first wire is a pure state, so the transmitted current must have a spin pointing in \emph{some} direction; this fact implies that no reduction in spin-polarization is possible in the coherent 1D limit. Other limitations to the SFET proposal have been simulated by Shafir \etal{}\cite{shafir04}

In this paper we discuss the general problem of coherent propagation of currents through quantum dots, focusing on the relationship of incident to exit spin-polarization of the currents. For the case of 2D ballistic chaotic scattering regions with strong spin-orbit interaction, we use random matrix theory to give analytic results for the expected values of spin-polarization in the exit currents. Once we can describe the ingoing current in terms of a density matrix, all of the conclusions will follow. Thus, the problem is generally broken into two parts: first, find the relevant input density matrix for the system of interest; second, propagate that density matrix to find the output currents and polarizations.
We choose the density matrix formalism to describe the input currents to the quantum dots, as it is flexible enough to describe any current in the noninteracting system.  As an important example, we describe how to construct the density matrices representing currents produced from potentials applied to (possibly spin-split) reservoirs. We go beyond this model and also consider injection of spin currents entangled between the two leads, finding larger fluctuations in this case. Similar work in a three-terminal geometry was considered in Ref.~\onlinecite{Adagideli09}. The case of unpolarized input currents was considered in Ref.~\onlinecite{Krich08}.

\section{Setup}

We consider a quantum dot attached to two ideal leads through quantum point contacts (QPCs). There are $\NL$, $\NR$ open spin-degenerate channels in the left, right QPCs, respectively, and we let $\Ntot=\NL+\NR$. We take a basis for the propagating states in the ideal leads normalized to unit flux in each channel, as usual. We consider noninteracting spin $1/2$ particles which are coherently scattered by the quantum dot, which we describe using an S-matrix. Given a density matrix $\win$ representing the current into the dot from the $\Ntot$ channels, the output current is described by density matrix $\wout=S \win S^\dagger$.

With $\Ntot$ open channels, the S-matrix $S$ can be represented by a $2\Ntot\times2\Ntot$ matrix of complex numbers. In systems with time reversal symmetry, it is convenient to consider $S$ to be a $\Ntot\times\Ntot$ matrix of $2\times2$ matrices. Any $2\times2$ matrix can be written as a linear combination of the four Pauli matrices, but it is convenient to consider the basis $\{\sigma^0,i\sigma^1,i\sigma^2,i\sigma^3\}$, where the $\sigma^i$ are the Pauli spin matrices. In this basis, a $2\times2$ matrix $q=q^0\sigma^0+i\vec{q}\cdot\vec\sigma$, with $q^0, \vec q\in\mathbb{C}$, which is also called a quaternion.\cite{Mehta04} Then $q$ is defined to have a complex conjugate $q^*=q^{0 *}\sigma^0+i\vec{q}^*\cdot\vec\sigma$, dual $q^R=q^0\sigma^0-i\vec{q}\cdot\vec\sigma$, and Hermitian conjugate $q^\dagger=q^{R*}$. The Hermitian conjugate is the same as the standard Hermitian conjugate of a complex matrix, but the complex conjugate is not the same.  For an S-matrix of quaternions, we define complex conjugate $(S^*)_{i j}=(S_{i j})^*$, dual $(S^R)_{i j}=(S_{j i})^R$, and Hermitian conjugate $S^\dagger=S^{R*}$. This representation is convenient because for time reversal invariant systems, $S=S^R$. The quaternion representation has the standard convention that $\Tr(S)=\sum_i S_{ii}^0$, which is half of the trace of the equivalent complex matrix.

\subsection{Constructing $w$ from chemical potentials}
Consider for the moment not two leads attached to the dot but $\Ntot$ leads, each with one open channel and connected to its own reservoir with adiabatic, reflectionless contacts.  Modeling the reservoirs as paramagnetic, each reservoir can be spin-split along its own quantization axis with each spin band separately in equilibrium, having its own chemical potential $\mu_{m}^\nu$, where $m\in\{1\dots\Ntot\}$ labels the channel and $\nu\in\{0,x,y,z\}$ indicates the charge and spin potentials.\cite{Veillette04,Tang02} There has been some confusion \cite{Sun08} on the consistency of defining this chemical potential, so we give an example. If reservoir $m$ is spin-split along axis $\hat x$, then $\mu_m^0$ is the average chemical potential in the reservoir, $2\mu_m^x$ is the chemical potential difference between spin-up and spin-down electrons quantized along $\hat x$, and $\mu_m^{y,z}=0$.  In general, if the quantization axis is $\hat n$ and the chemical potential difference along that axis is $2\mu^s$, then $\mu^i=\mu^s(\hat n \cdot \hat i)$. Such spin-split chemical potentials can be realized, for example, by optical excitation in heterostructures, in an environment with inelastic relaxation much faster than spin relaxation.\cite{Veillette04,Malajovich00,Kikkawa99}

We assume the leads have negligible spin-orbit coupling and spin relaxation, so there is a well-defined spin current in the leads. In the absence of inelastic processes, we can consider the current carried by particles with energy $\epsilon$. For simplicity, we assume the number of open channels does not vary over the range of $\epsilon$ considered here. Then the particle-currents flowing in from each channel are represented by the quaternion density matrix
\begin{align}
  \tilde\win_{nm}(\epsilon)=\delta_{nm}[ f(\epsilon-\mu^0_n)-\vec\sigma\cdot\vec\mu^s_n f^\prime(\epsilon-\mu^0_n)],
\end{align}
where $f(\epsilon)$ is the Fermi function at temperature $T$, and we assume that $\mu^s_n<\max(T,\Delta)$, where $\Delta$ is the mean orbital level spacing in the quantum dot without leads attached, and the prime indicates the derivative with respect to $\epsilon$.

The charge current in the $n^\text{th}$ channel of particles with energy $\epsilon$ is
\begin{align}
  j^0_n(\epsilon)=2\Tr\{P_n[\tilde\win(\epsilon)-\tildewout(\epsilon)]\} \frac{e}{h},
\end{align}
where $-e$ is the electron charge, $h$ is Planck's constant, and $P_n$ is the projection matrix onto the $n^\text{th}$ channel (i.e., $(P_n)_{ab}=\delta_{an}\delta_{bn}$). Similarly, the spin-current in the $n^\text{th}$ channel is
\begin{align}\label{eq:cur_def}
  j^i_n(\epsilon)=2\Tr\lbrace P_n\sigma^i [\tilde\win(\epsilon)-\tildewout(\epsilon)]\rbrace\frac{e}{2\pi}.
\end{align}
We choose units in which $e=h=2\pi$, so Eq.~\ref{eq:cur_def} can describe both charge and spin currents if we let $\sigma^0$ be the identity.  

The currents are the physical objects in the system, and we note that the currents are unaffected by adding any multiple of the identity to $\tilde\win(\epsilon)$, since $\tildewout=S\tilde\win \Sdag$ and $S$ is unitary. We can thus use the density matrix to represent the currents, but we do not need to maintain $\Tr \win=1$ or even that $\win$ has positive eigenvalues. In the case where there are only two leads, we can subtract $f(\epsilon-\mu_2^0)$ from $\win$, giving
\begin{widetext}
\begin{align}
  {\win}(\epsilon)&=\begin{pmatrix}
    \left[f\left(\epsilon-\mu_1^0\right)-f\left(\epsilon-\mu_2^0\right)\right] -f^\prime\left(\epsilon-\mu_1^0\right) \vec\sigma\cdot \vec\mu^s_1\\
    &-f^\prime\left(\epsilon-\mu_2^0\right) \vec\sigma\cdot \vec\mu^s_2
  \end{pmatrix}\nonumber\\
  &\approx \begin{pmatrix}
    -f^\prime\left(\epsilon-\mu^0\right)\delta\mu^0 -f^\prime\left(\epsilon-\mu_1^0\right) \vec\sigma\cdot \vec\mu^s_1\\
    &-f^\prime\left(\epsilon-\mu_2^0\right) \vec\sigma\cdot \vec\mu^s_2
  \end{pmatrix},
\end{align}
\end{widetext}
where $\mu^0=(\mu_1^0+\mu_2^0)/2$ and $\delta\mu^0=\mu_1^0-\mu_2^0$. Note that if $\delta\mu^0=0$ then the average chemical potential in both leads is the same, so no net charge flows and $\win$ is traceless.

If we consider an energy range in which the S-matrix does not vary (i.e., the linear response regime,\cite{datta} where $\delta\mu^\nu<\{T,\Delta\}$), then we can represent the currents by integrating over energy in the density matrix, giving
\begin{align}\label{eq:win}
  {\win}= \begin{pmatrix}
  \delta\mu^0+\vec\sigma\cdot\vec\mu_1^s\\
  &\vec\sigma\cdot\vec\mu_2^s\\
  \end{pmatrix}
\end{align}
and
\begin{align}\label{eq:j_def}
  j_n^\nu=2\Tr[P_n\sigma^\nu({\win}-{\wout})].
\end{align}

Spin-polarized injection from ferromagnetic contacts does not immediately map onto the chemical potential formalism. It is clear that if a ferromagnet is in equilibrium with a wire, connected by adiabatic contacts, it will not produce a spin current in the wire, since adiabaticity requires that the lowest energy levels remain filled. For practical injection of spin-polarized currents from a ferromagnet to a normal metal system, a tunnel barrier at the contact is the most common form of non-adiabaticity.\cite{Rashba00,Lou07}

We can consider a situation where the ferromagnet injects into a semiconductor, which serves as the reservoir for a wire connected to our quantum dot.  If we consider the case where the semiconductor has an energy relaxation time $\tau_e$ much shorter than the spin relaxation time $\tau_s$, then the spin-polarized current injected from the ferromagnet into the reservoir can relax to two independent distributions with a spin-split chemical potential. This is the same assumption used for optical excitation of spin-split chemical potentials. We can then use the formulation in terms of potentials as described above.

The tunnel barrier at the ferromagnet introduces a second complication, as it implies that the ingoing current in the wire contains particles injected directly from the reservoir and also particles reflected from the scattering region and reflected back from the barrier.  The input density matrix thus needs to be determined self-consistently, including the effects of both reflections. Such effects can be included systematically, by using the Poisson kernel \cite{Beenakker97} rather than the circular ensemble described below and also including the TRS-breaking effects of the ferromagnetic scattering. For a sufficiently large reservoir in the semiconductor, this reflection can represent a small perturbation to the input currents, and the procedure described below will be a good approximation.

\subsection{Connection to spin conductances}

We can write a generalized B\"uttiker-type conductance equation \cite{Buttiker86}
\begin{align}\label{eq:JGmu}
  j_l^\nu=\sum_{k,\rho}G_{lk}^{\nu\rho} \mu_k^{\rho}-2 M_l\mu_l^\nu,
\end{align}
where $G_{lk}^{\nu\rho}$ is the conductance from lead $k$ to lead $l$ and spin $\rho$ to $\nu$ and $2 M_l$ is the number of modes, including spin, in lead $l$. The absence of equilibrium charge or spin currents (since there is no spin-orbit coupling in the leads) implies
\begin{align}
  \sum_{k}G_{lk}^{\nu 0}-2M_l=0.
\end{align}
Further, the conservation of charge current implies that 
\begin{align} 
  \sum_{l} G_{lk}^{0 \nu}=2M_k\delta_{\nu 0}.
\end{align}

Specializing to the case of two leads with $\NL$ and $\NR$ modes in the left and right leads with potentials $\mu^\nu_L$, $\mu^\nu_R$, respectively, we can express $G^{\nu\rho}_{lk}$ simply in terms of the S-matrix. Setting $\mu^\nu_R=0$ and $\mu^\nu_L=\delta_{\nu \alpha}$, Eq.~\ref{eq:JGmu} gives
\begin{align}
  j_R^\nu=G_{RL}^{\nu\alpha}.
\end{align}
Eq.~\ref{eq:win} says that $\win=\left(\begin{smallmatrix} \sigma^\alpha\openone_\NL\\ &0_\NR\end{smallmatrix}\right) =\sigma^\alpha P_L$, and by Eq.~\ref{eq:j_def} we have $j_R^\nu=2\Tr(\sigma^\nu P_R S \sigma^\alpha P_L \Sdag)=G_{RL}^{\nu\alpha}$.
Similarly, $G^{\nu\alpha}_{RR}=2\Tr(\sigma^\nu P_R S \sigma^\alpha P_R \Sdag)$. If the system is time reversal invariant, then $S=S^R$, which imposes some relations between the different conductance matrix elements. Since $\Tr(A^R)=\Tr(A)$, we have the Onsager-like relations
\begin{align}
  G_{lk}^{\nu\rho}=h^\nu h^\rho G_{kl}^{\rho \nu},
\end{align}
for $k,l=R,L$ where $h^\nu=(1,-1,-1,-1)$.

We thus see that we can express all of the $G_{ij}^{\nu\rho}$ in terms of traces over appropriate density matrices multiplying S-matrices.
We will consider the current in the right lead associated with the input density matrix $\win$, defined as
\begin{align}\label{eq:g^nu_def}
  \jr\nu\equiv 2\Tr[\sigma^\nu P_R(S\win \Sdag-\win)],
\end{align}
which is proportional to the outgoing current in the right lead after injection represented by $\win$, where the sign is chosen so that outgoing currents to the right are positive. 

\subsection{Purity of \win\label{sec:purity}}
We will see that the coherence properties of the currents are important, so it is interesting to consider when $\win$ represents a pure state.  Ordinarily, density matrices are defined (with quaternion trace convention) so $2\Tr \rho=1$, and $\rho$ is pure if $\rho^2=\rho$. In our open system, normalization is a choice, and we set $2\Tr w=\Ta$, where $\Ta$ gives the total current incident on the dot. We can also add any multiple of the identity to $w$ without affecting the physical currents. Taking both these factors into account, $w$ represents a pure state only if there is a real number $\alpha$ such that
\begin{align}
  \left(\frac{w-\alpha\openone}{2\Tr(w-\alpha\openone)}\right)^2=  \frac{w-\alpha\openone}{2\Tr(w-\alpha\openone)}.
\end{align}
This condition implies that the $\Ntot\times\Ntot$ quaternion matrix $w$ represents a pure state only if
\begin{enumerate}
  \item $w^2=\Ta w $,
  \item $w^2=\frac{-\Ta}{2\Ntot-1}w$, or
  \item $w$ is invertible and $\exists\alpha\in\mathcal{R}$ such that \\ $w^{-1}=\frac{w-[\Ta-2\alpha(\Ntot-1)]\openone}{-\alpha[\Ta-\alpha(2\Ntot-1)]}$.
\end{enumerate}

\section{Random matrix theory}
Though for any particular quantum dot it is difficult to determine the full scattering matrix exactly, if there is a small number of open channels in the leads connected to the dot, mesoscopic fluctuations should produce an appreciable spin polarization in the exit current. We can understand this by considering that the current from one of the input channels has some probability to exit into each of the $\NR$ exit channels after undergoing some spin rotation. In the chaotic strong spin-orbit limit, there is no correlation between the entry and transmitted spin polarizations. Though on average the transmitted spin polarization is zero, in any particular case there will still be some residual polarization in some direction in the exit lead. When there is only a small number of channels in the entrance and exit, these residual polarizations can be large. We will find the root mean square spin currents in the right lead by averaging over the ensemble of coherent cavities with strong spin-orbit coupling. These fluctuations are due to mesoscopic interference effects inside the quantum dots.
We are primarily interested in the time-reversal invariant case, but we will present results valid with and without TRS.

We consider coherent elastic scattering of noninteracting electrons with no spin-relaxation in the leads. We consider the chaotic limit for the quantum dot, in which the electron dwell time $\tau_d=2\pi\hbar/\Ntot \Delta$ is much longer than the Thouless time $\tau_\text{Th}=L_d/v_F$, where $L_d$ is a typical linear distance across the dot, $v_F$ is the Fermi velocity, and $\Delta=2\pi\hbar^2/m A$ is the mean orbital level spacing in the quantum dot, with $m$ the effective mass and $A$ the area of the dot. We further assume the strong spin-orbit limit, where the spin-orbit time $\tau_\text{so}$ is much less than $\tau_d$. We assume that all of the channels have perfect coupling into the quantum dot. 

We are interested in the properties of the current in the right lead. For an input density matrix $\win$, in addition to $\jr\nu$, we define the outgoing current
\begin{align}
  \g^\nu  &=2\Tr\left(\sigma^\nu P_R S \win \Sdag\right),
\end{align}
and the current due only to the input state
\begin{align}
  \Jr\nu=2\Tr(\sigma^\nu P_R \win)
\end{align}
so $\jr{\nu}=\g^\nu-\Jr{\nu}$. The charge current is $\jr{0}$ and the spin current is $\vecjr$. We define $\jr{s}=\abs{\vecjr}$. The polarization of the current in the right lead is $\vecp=\vecjr/\jr{0}$.
A small number of parameters of the input current are sufficient to describe the effects of any $\win$ in a two-terminal configuration. In particular, we define
\begin{align}
  \Ta&=2\Tr\win\\
  C&=2\Tr({\win}^2)\\
  D^\nu&=2\Tr(\sigma^\nu P_R{\win}^R)=(\Jr\nu)^R\\
  E^\nu&=2\Tr(\sigma^\nu P_R{\win}^R{\win}^R)\\
  F^\nu&=2\Tr(\sigma^\nu P_R{\win}^R P_R\sigma^\nu {\win}^R),
\end{align}
where superscript $R$ is the quaternion dual, $\Ta$ is the total flux incident on the dot, $C$ is a measure of the coherence of the current, $D^\nu$ gives the incident charge/spin current from the right lead, $E^\nu$ and $F^\nu$ are more measures of coherence. By adding a multiple of $\openone$ to $\win$, we can choose $D^0=2\Tr(P_R\win)=0$, and all results below assume this choice. Note that if current is incident only from the left lead, then $D^\nu=E^\nu=F^\nu=0$.

We take averages over the uniform ensemble of all S-matrices in the strong spin-orbit limit, either with TRS (called the circular symplectic ensemble -- CSE) or without TRS (called the circular unitary ensemble -- CUE).\cite{Mehta04,Beenakker97}  Such averaging is readily performed experimentally by small changes of the shape of a quantum dot \cite{Zumbuhl02}; the root mean square (rms) fluctuations also give a typical value to be expected for any one chaotic dot. An external magnetic field can easily break TRS, moving between these ensembles. A convenient formalism for performing such averages was worked out by Brouwer and Beenakker.\cite{Brouwer96}  From that work, we need two averages. In the quaternion representation, for $f_1=\Tr(A S B S^\dagger)$ for $A,B$ constant $\Ntot \times \Ntot$ quaternion matrices,
\begin{align}\label{eq:f1}
  \av{f_1}_\CSE&=\frac{1}{2\Ntot-1}[2\Tr A\Tr B-\Tr(A^R B)]\\
  \av{f_1}_\CUE&=\frac{1}{\Ntot}\Tr A \Tr B.
\end{align}

\newcommand{\Lu}{\Lambda_U}
\newcommand{\Ls}{\Lambda_S}
\newcommand{\ds}{\delta_S}
\newcommand{\Lb}{\Lambda}
The other average we need is of $f_2=\Tr(A S B \Sdag)\Tr(A S B \Sdag)$ for $A$, $B$ constant $\Ntot\times\Ntot$ quaternion matrices. We find \cite{Brouwer96}
\begin{widetext}
\begin{align}\label{eq:f3} 
  \av{f_2}_\CSE=&\frac{1}{2\Ls}\bigg(\lbrace\Ntot-1\rbrace
  \Big\lbrace 8[\Tr A]^2[\Tr B]^2+2\Tr[A^2]\Tr[B^2]+4[\Tr(A B^R)]^2
  -8\Tr[A]\Tr[B]\Tr[A B^R]-2\Tr[A A B^R B^R]\Big\rbrace\nonumber\\
  &-\Big\lbrace 2[\Tr A]^2\Tr[B^2]+\Tr[A^2][\Tr B]^2-4\Tr[A]\Tr[A^R B^2]-4\Tr[B]\Tr[A^2 B^R]\nonumber\\
  &+4\Tr[A]\Tr[B]\Tr[A B^R]
  +\Tr[A B^R A B^R]+\Tr[A A B^R B^R]\Big\rbrace\bigg)\\
  \av{f_2}_\CUE=&\frac{1}{\Lu}\left[4\Ntot(\Tr A)^2 (\Tr B)^2+\Ntot\Tr(A^2)\Tr(B^2) -\Tr(A^2)(\Tr B)^2- (\Tr A)^2 \Tr(B^2)\right],\label{eq:f3U}
\end{align}
\end{widetext}
where $\Ls=\Ntot(2\Ntot-1)(2\Ntot-3)$ and $\Lu=\Ntot(4\Ntot^2-1)$.

We consider the mean and fluctuations of $\jr\nu$. Using Eq.~\ref{eq:f1},
\begin{align}
  \av{\jr\nu}&=\av{\g^\nu}-\Jr{\nu}\label{eq:jr_av}\\
  \av{\g^\nu}&=\delta_{\nu0}\frac{2\Ta\NR}{2\Ntot-\ds} - \ds\frac{D^\nu}{2\Ntot-1},\label{eq:g_nu_av}
\end{align}
where $\ds=1$ for averages over the CSE and $\ds=0$ for averages over the CUE.
The relevant fluctuations to study are of $\Delta\jr\nu=\jr\nu-\av{\jr\nu}$, which satisfy
\begin{alignat}{10}
  \av{\left(\Delta\jr\nu\right)^2}&=\av{{\g^\nu}^2}-\av{\g^\nu}^2
\end{alignat}
Using Eqs.~\ref{eq:f3} and \ref{eq:f3U}, we find
\begin{align}\label{eq:g_nu^2}
  \av{{\g^\nu}^2}=\frac{1}{\Lb}&\left\lbrace \NR\delta_{\nu 0}\big[4\NR\Ta^2(\Ntot-\ds) - 2\NR C+4 \ds E^0\right] \nonumber\\
   &-\NR\Ta^2 + (\Ntot-\ds)(2\NR C+2\ds{D^\nu}^2)\\\nonumber
   &-\ds[E^0(2\Ntot-1)-F^\nu]\big\rbrace,
\end{align}
where $\Lb=\Ls$, $\Lu$ for the CSE, CUE, respectively.
We note that $\av{{\g^\nu}^2}_\CUE$ does not depend on $D$, $E$, or $F$. Combining this result with Eq.~\ref{eq:g_nu_av},
\begin{widetext}
\begin{align}\label{eq:fluct} 
  \av{{\g^\nu}^2}-\av{\g^\nu}^2=\frac{1}{\Lb}\Big\lbrace &\NR\delta_{\nu 0}\left[\frac{\NR\Ta^2(1+\ds)}{\Ntot-\ds/2} - 2\NR C+4 \ds E^0\right] \\\nonumber
  &+ 2\NR C(\Ntot-\ds)
  -\NR\Ta^2+\ds[(D^\nu)^2\frac{2\Ntot^2-3\Ntot+2}{2\Ntot-1} - E^0(2\Ntot-1) - F^\nu]\Big\rbrace
\end{align}
\end{widetext}

Eq.~\ref{eq:fluct} is the main result of this work, and we will now look at its implications in some special cases. First, an arbitrarily polarized current incident from the left lead, as can be readily created by optical methods. Second, a pure spin current uniformly distributed between the leads. Third, a pure state pure spin current, with entanglement between the currents incident from each lead.

\subsection*{Case 1: Spin-polarized current}
For any current incident exclusively from the left, the total current $\Ta$ and the parameter $C$ are sufficient to describe mean and rms currents in the right lead. We consider the input current represented by
\begin{align}\label{eq:w1}
  \win_1= \frac{1}{2\NL}
  \begin{pmatrix}
     \openone_\NL(\Ta\sigma_0+\vec s\cdot \vec \sigma)\\
     &0_\NR
  \end{pmatrix}
\end{align}
where $\vec s$ is the polarization magnitude and direction of the input spin current. Note that $\Ta$ can be positive, negative, or zero, depending on the direction of the charge current through the device. For $\abs{\vec s}=\abs{t}$, the current is fully polarized. 

For the density matrix of Eq.~\ref{eq:w1}, $C=(\Ta^2+s^2)/2\NL$, and $D=E=F=0$. Applying Eq.~\ref{eq:g_nu_av}, the mean spin current in the right lead is zero and the average charge current is $\av{\jr{0}}=2\Ta\NR/(2\Ntot-\ds)$. The reduction of $\av{\jr{0}}$ as TRS is broken ($\ds\rightarrow0$) is the signature of weak antilocalization. \cite{Beenakker97,Bergmann82,Chakravarty86} 
The rms spin current in the right lead is
\begin{align} 
  \av{{\jr{s}}^2}=& 3\frac{\NR[(\NR-\ds)\Ta^2+(\Ntot-\ds)s^2]}{\NL \Lb}\label{eq:js2_w1}.
\end{align}
The fluctuations in the charge current are
\begin{align} 
  \av{{\Delta\jr{0}}^2}=\frac{\NR}{\Lb} \left\lbrace[4\NR\NL -\ds(4\NR-\frac{1}{\NL})]\Ta^2+(1-\frac{\ds}{\NL})s^2\right\rbrace \nonumber
\end{align}

\begin{figure}
  \includegraphics[width=\figwidth]{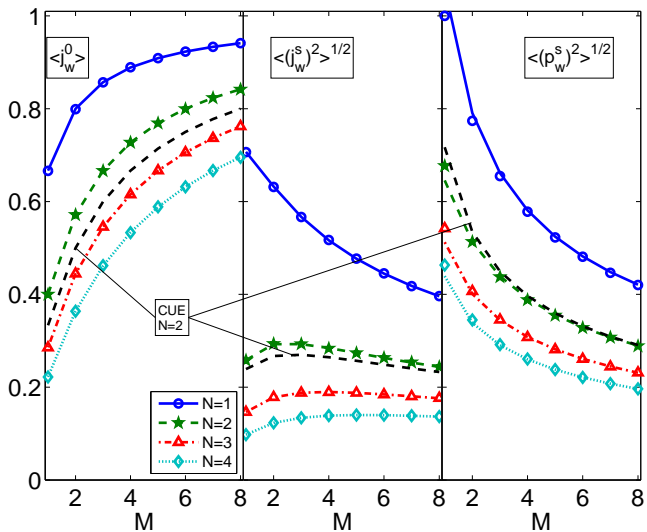}
  \caption{\label{fig:w1} (color online) For the fully spin-polarized current represented by Eq.~\ref{eq:w1} with $\Ta=s=1$ and time reversal symmetry, comparison of numerical (symbols) and analytical (lines) results for the mean charge current (left), rms spin current (middle) and rms spin polarization (right) in the exit lead, where $\NR$ ($\NL$) is the number of channels in the exit (entrance) lead. An average over 50 000 S-matrices from the CSE was performed for each data point. The lines are from Eqs.~\ref{eq:jr_av}, \ref{eq:js2_w1}, and \ref{eq:p2_w1}. The right panel shows that the expected spin polarization in the exit lead is still appreciable, even for several open modes in each of the leads. Also shown are the equivalent CUE results with $\NL=2$, showing that the rms spin polarization is nearly unchanged by breaking TRS in this case.}
\end{figure}

In the case of an unpolarized charge current ($s=0$) with TRS, spin current in the exit lead is forbidden when $\NR=1$ due to the combined effects of time reversal symmetry and unitarity,\cite{Krich08,Kiselev05,Zhai05} as can be seen in Eq.~\ref{eq:js2_w1}.
We can consider a pure spin current incident from the left by setting $\Ta=0$.  In that case, we see that
\begin{align} \label{eq:j02_w1}
  \av{{\Delta\jr{0}}^2}= \frac{\NR(\NL-\ds)s^2}{\NL\Lb},
\end{align}
showing the scale of charge currents produced from the pure spin current. Similar effects have recently been proposed to measure the spin conductance in a three-terminal geometry.\cite{Adagideli09}
We note that $\langle{\jr{0}}^2\rangle_\CSE=0$ if $\NL=1$, showing that a pure spin-current incident from a single channel cannot produce a net charge current in the other channels. This is the time reversed statement of the theorem that with TRS a charge current cannot produce a spin-polarized current when $\NR=1$.

We can further consider the spin-polarization of the exit current, $\vecp=\vecjr/\jr{0}$.  It is clear that $\av{\vecp}=0$, just as $\langle\vecjr\rangle=0$, but there is some rms spin polarization of the exit current. If we approximate $\av{\p^2}\approx\av{{\jr{s}}^2}/\av{\jr{0}}^2$, we can use the above results to find
\begin{align}\label{eq:p2_w1}
    \av{\p^2}\approx 3 (\Ntot-\ds/2)^2\frac{\Ta^2(\NR-\ds)+s^2(\Ntot-\ds)} {\Lb \Ta^2\NR\NL}.
\end{align}
To test this approximation, we found $\av{\p^2}$ by numerically averaging over the CSE. Matrices drawn from the CSE were chosen by diagonalizing matrices from the Gaussian Unitary Ensemble, as described in Ref.~\onlinecite{Krich08}. Results are shown in Fig.~\ref{fig:w1} for the case $\Ta=s=1$, and it is clear that Eq.~\ref{eq:p2_w1} agrees very well with the numerical results (right panel). 
The case shown in the figure is the relevant one for the Schliemann-Egues-Loss SFET, in which a fully polarized spin current is incident from one lead. In the off state, which relies on large spin-orbit coupling, the spin polarization in the exit lead is supposed to be zero. We see in Fig.~\ref{fig:w1} that even for several open channels in each lead, we expect to find an appreciable spin polarization in the output, limiting off-state function.

\subsection*{Case 2: Pure spin current from both leads}

\begin{figure}
  \includegraphics[width=\figwidth]{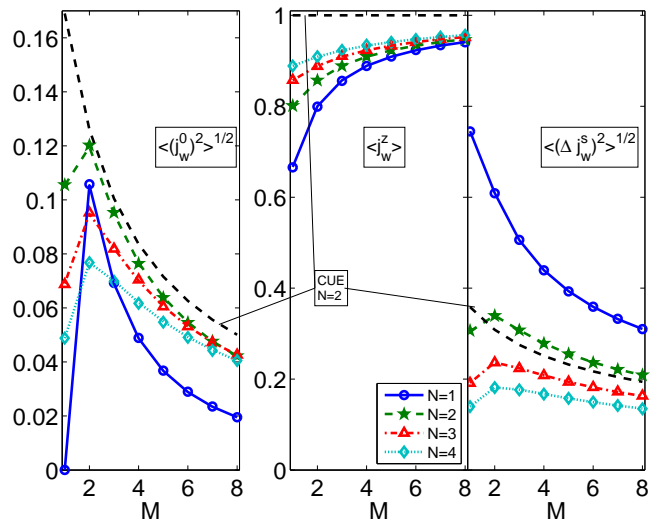}
  \caption{\label{fig:w2} (color online) For the pure spin current represented by Eq.~\ref{eq:w2}, comparison of numerical (symbols) and analytical (lines) results for the rms charge current (left), mean spin current (middle) and rms spin current fluctuations (right) in the exit lead, where $\NR$ ($\NL$) is the number of channels in the exit (entrance) lead. An average over 50 000 S-matrices from the CSE was performed for each data point.  The lines are from Eqs.~\ref{eq:jr02_w2}--
  \ref{eq:j2z_w2}. The left panel shows that this pure spin current should still be expected to produce significant charge currents, with a nonmonotonic dependence on the number of open channels $\NL$ and $\NR$. Also shown are the equivalent CUE results with $\NL=2$.}
\end{figure}

We consider a pure spin current incident from both leads, represented by the density matrix
\begin{align}\label{eq:w2}
  \win_2=
  \begin{pmatrix}
    \openone_\NL \frac{\sigma_z}{2\NL}\\
    &-\openone_\NR \frac{\sigma_z}{2\NR}
  \end{pmatrix}.
\end{align}
This density matrix represents a spin current of $+\hat{z}$ incident from the left and a spin current of $-\hat{z}$ incident from the right, which together are an incident pure spin current from left to right with polarization $+\hat z$. In this case, $\Ta=0$, $C=\Ntot/2\NR\NL$, $D^\nu=(0,0,0,1)$, $E^\nu=(1/2\NR,0,0,0)$, and $F^\nu=(1,-1,-1,1)/2\NR$. Though the mean value of the charge current is zero, since it is as likely for the charge current to flow in as out, the spin current can produce a mean square charge current
\begin{align}\label{eq:jr02_w2}
  \av{{\jr{0}}^2}=\frac{\Ntot}{\Lb} \left[1 - \ds\frac{\NL^2+\NR^2}{\NR\NL\Ntot}  \right].
\end{align}
We note that when $\NR=\NL=1$, $\av{j_0^2}_\CSE=0$, showing that no charge current can be produced. This result is another implication of the theorem that, with time reversal symmetry, a spin current incident in one channel cannot produce a charge current, combined with a simpler result that coherence and time reversal symmetry forbid spin-to-charge reflection in a single channel.

The spin current in the right lead is a combination of the incident spin current, the reflected spin current from the right and the transmitted spin current from the left.  Together, these give a mean spin current of
\begin{align}\label{eq:jr_av_w2}
  \av{\jr{i}}=(0,0,1-\ds\frac{1}{2\Ntot-1}).
\end{align}
Thus, with TRS, the spin current in the right lead is, on average, reduced from 1. In the case $\NR=\NL=1$, this reduction removes 1/3 of the spin current that began in the lead.

The fluctuations around the mean are
\begin{align}
  \av{{\Delta\jr{{x,y}}}^2}&=
    \frac{\Ntot(\Ntot-\ds)\left(1-\ds\frac{\NL}{\NR\Ntot}\right)}{\NL\Lb }\label{eq:j2xy_w2}\\
  \av{{\Delta\jr{z}}^2}&=\frac{\Ntot}{(\Ntot-\ds/2)\NL\Lb}\Big\lbrace\Ntot^2 +\label{eq:j2z_w2}\\\nonumber
  \ds \Big[&\frac{(\NR-1)\Ntot^2}{\NR}-\Ntot(\NR+2-\frac{1}{2\NR}) +\frac{3\NR}{2}+\frac{\NL}{\Ntot}\Big]\Big\rbrace
\end{align}

These results, along with confirming numerical simulations, are shown in Fig.~\ref{fig:w2}.

\subsection*{Case 3: Pure state pure spin current}
We consider entanglement between the currents in the two leads, which is beyond the standard chemical potential formulation of transport.  In particular, consider a pure state spin current entangled between both leads, rather than the mixed state spin current of case 2.  With $\NR=\NL=1$, we consider
\begin{align}\label{eq:w3}
  \win_3=\frac{1}{2}
  \begin{pmatrix}
    \sigma_z        & \sigma_x+i\sigma_y\\
    \sigma_x-i\sigma_y & -\sigma_z
  \end{pmatrix}
\end{align}
This state has, as in case 2, a pure spin current $+\hat{z}$ incident from the left and a pure spin current $-\hat{z}$ incident from the right, but the off-diagonal terms of $\win_3$ indicate that the currents are entangled.
The density matrix formalism easily allows consideration of such off-diagonal correlations between the channel currents. 
The entanglement could be produced by passing a current through a beamsplitter produced from quantum dots\cite{Oliver02,Saraga03,Kolli09}, feeding into the two channels or from spin injection by optical orientation using entangled photons. The density matrix $\win_3$ represents a pure state by condition 3 of section \ref{sec:purity} with $\alpha=-1/2$.

In this scenario, $\Ta=0$, $C=3$, $D^\nu=(0,0,0,1)$, $E^\nu=(3/2,0,0,1)$, and $F^\nu=(1,-1,-1,1)/2$.  This should be compared with case 2 in the $\NR=\NL=1$ limit, which is the same except $C=1$ and $E^\nu=(1/2,0,0,0)$.

The most significant difference from case 2 is that coherence between the channels allows a charge current to be produced, even when $\NR=\NL=1$ with TRS. There is still no mean charge current, but the rms fluctuations are $\langle{\jr{0}}^2\rangle_\CSE^{1/2}=0.41$, $\langle{\jr{0}}^2\rangle_\CUE^{1/2}=0.45$. This rms charge current is much larger than the results of case 2, even away from $\NR=\NL=1$, (see Fig.~\ref{fig:w2}, left, and Eq.~\ref{eq:jr02_w2}) indicating that the entangled spin current is better able to couple into charge current than is the incoherent spin current. We have normalized $\win_3$ to have $\av{\jr i}=(0,0,1-\ds/3)$, as in case 2.  We find fluctuations around the mean of $\langle{\Delta\jr{i}}^2\rangle_\CSE^{1/2}=(0.58,0.58,0.62)$, $\langle{\Delta\jr{i}}^2\rangle_\CUE^{1/2}=(0.63,0.63,0.63)$. With TRS, the fluctuations are larger along the polarization axis, but not markedly so.  The total spin polarization fluctuations are $\av{{\Delta\jr{s}}^2}_\CSE^{1/2}=1.03$, $\av{{\Delta\jr{s}}^2}_\CUE^{1/2}=1.10$ which is larger than the mean current and equal in scale to the input current $\Jr{s}$, showing that coherence between the channels significantly enhances the mesoscopic fluctuations; this should be compared with Fig.~\ref{fig:w2} (right panel). Such large fluctuations entail a significant loss of knowledge of the quantization axis of the spin current, so the initially $z$-polarized current can exit polarized in many directions.

\section{Discussion}
Mesoscopic fluctuations of spin current on passing through a chaotic ballistic quantum dot can produce large fluctuations in spin-polarization, charge currents from pure spin currents, and spin currents from charge currents.\cite{Krich08}
These predictions for mean and rms currents will be modified by dephasing and effects of the energy dependence of the S-matrix. Dephasing processes can be readily added to this model using the third lead method,\cite{Buttiker88,Brouwer95,baranger95} as detailed in Ref.~\onlinecite{Krich08}. Dephasing generally reduces the fluctuations in charge and spin currents and also removes the symmetry that forbids charge or spin currents at certain values of $\NR$ and $\NL$ with TRS.

If the ingoing current contains particles with energies varying over a large enough range, the energy dependence of the S-matrix must be considered as well. The S-matrix is generally correlated on the energy scale of the level broadening of the quantum dot eigenstates, approximately $\Delta^\prime=\Delta\Ntot/2+\gamma_\phi/2$, where $\gamma_\phi$ is the dephasing rate.\cite{Huibers98,Krich08} If the incident particles have energies that differ by a large amount compared to the level broadening $\Delta^\prime$, as can happen at sufficiently large temperatures or $\delta\mu^\nu$, then the mesoscopic fluctuations are suppressed, as there are effectively more open channels for particles passing through the dot.

Mesoscopic fluctuations producing spin polarized exit currents could be important for operation of a Schliemann-Egues-Loss SFET. To avoid this impact on the off-state polarization, such a device should have many scattering regions in parallel or operate in a regime with sufficiently large temperature, bias, or dephasing so as to reduce these mesoscopic effects.

\begin{acknowledgments}
  We acknowledge useful discussions and a careful reading of this manuscript by Bert Halperin. We acknowledge helpful conversations with Caio Lewenkopf, Emmanuel Rashba, Ilya Finkler, and Ari Turner. This work was supported in part by the Fannie and John Hertz Foundation and NSF grant PHY-0646094.
\end{acknowledgments}

\bibliography{SFET,CubicDressRefs}

\begin{thebibliography}{38}
\expandafter\ifx\csname natexlab\endcsname\relax\def\natexlab#1{#1}\fi
\expandafter\ifx\csname bibnamefont\endcsname\relax
  \def\bibnamefont#1{#1}\fi
\expandafter\ifx\csname bibfnamefont\endcsname\relax
  \def\bibfnamefont#1{#1}\fi
\expandafter\ifx\csname citenamefont\endcsname\relax
  \def\citenamefont#1{#1}\fi
\expandafter\ifx\csname url\endcsname\relax
  \def\url#1{\texttt{#1}}\fi
\expandafter\ifx\csname urlprefix\endcsname\relax\def\urlprefix{URL }\fi
\providecommand{\bibinfo}[2]{#2}
\providecommand{\eprint}[2][]{\url{#2}}

\bibitem[{\citenamefont{Kikkawa and Awschalom}(1999)}]{Kikkawa99}
\bibinfo{author}{\bibfnamefont{J.~M.} \bibnamefont{Kikkawa}} \bibnamefont{and}
  \bibinfo{author}{\bibfnamefont{D.~D.} \bibnamefont{Awschalom}},
  \bibinfo{journal}{Nature} \textbf{\bibinfo{volume}{397}},
  \bibinfo{pages}{139} (\bibinfo{year}{1999}).

\bibitem[{\citenamefont{Stich et~al.}(2007)\citenamefont{Stich, Zhou, Korn
  et~al.}}]{Stich07}
\bibinfo{author}{\bibfnamefont{D.}~\bibnamefont{Stich}},
  \bibinfo{author}{\bibfnamefont{J.}~\bibnamefont{Zhou}},
  \bibinfo{author}{\bibfnamefont{T.}~\bibnamefont{Korn}},
  \bibinfo{author}{\bibfnamefont{R.}~\bibnamefont{Schulz}},
  \bibinfo{author}{\bibfnamefont{D.}~\bibnamefont{Schuh}},
  \bibinfo{author}{\bibfnamefont{W.}~\bibnamefont{Wegscheider}},
  \bibinfo{author}{\bibfnamefont{M.~W.} \bibnamefont{Wu}}, \bibnamefont{and}
  \bibinfo{author}{\bibfnamefont{C.}~\bibnamefont{Sch\"{u}ller}},
  \bibinfo{journal}{Phys. Rev. Lett.} \textbf{\bibinfo{volume}{98}},
  \bibinfo{eid}{176401} (\bibinfo{year}{2007}).

\bibitem[{\citenamefont{Eto et~al.}(2005)\citenamefont{Eto, Hayashi, and
  Kurotani}}]{Eto05}
\bibinfo{author}{\bibfnamefont{M.}~\bibnamefont{Eto}},
  \bibinfo{author}{\bibfnamefont{T.}~\bibnamefont{Hayashi}}, \bibnamefont{and}
  \bibinfo{author}{\bibfnamefont{Y.}~\bibnamefont{Kurotani}},
  \bibinfo{journal}{J. Phys. Soc. Jpn.} \textbf{\bibinfo{volume}{74}},
  \bibinfo{pages}{1934} (\bibinfo{year}{2005}).

\bibitem[{\citenamefont{Bardarson et~al.}(2007)\citenamefont{Bardarson,
  Adagideli, and Jacquod}}]{Bardarson07}
\bibinfo{author}{\bibfnamefont{J.~H.} \bibnamefont{Bardarson}},
  \bibinfo{author}{\bibfnamefont{I.}~\bibnamefont{Adagideli}},
  \bibnamefont{and} \bibinfo{author}{\bibfnamefont{P.}~\bibnamefont{Jacquod}},
  \bibinfo{journal}{Phys. Rev. Lett.} \textbf{\bibinfo{volume}{98}},
  \bibinfo{eid}{196601} (\bibinfo{year}{2007}).

\bibitem[{\citenamefont{Krich and Halperin}(2008)}]{Krich08}
\bibinfo{author}{\bibfnamefont{J.~J.} \bibnamefont{Krich}} \bibnamefont{and}
  \bibinfo{author}{\bibfnamefont{B.~I.} \bibnamefont{Halperin}},
  \bibinfo{journal}{Phys. Rev. B} \textbf{\bibinfo{volume}{78}},
  \bibinfo{eid}{035338} (\bibinfo{year}{2008}).

\bibitem[{\citenamefont{Lou et~al.}(2007)\citenamefont{Lou, Adelmann, Crooker
  et~al.}}]{Lou07}
\bibinfo{author}{\bibfnamefont{X.}~\bibnamefont{Lou}},
  \bibinfo{author}{\bibfnamefont{C.}~\bibnamefont{Adelmann}},
  \bibinfo{author}{\bibfnamefont{S.~A.} \bibnamefont{Crooker}},
  \bibinfo{author}{\bibfnamefont{E.~S.} \bibnamefont{Garlid}},
  \bibinfo{author}{\bibfnamefont{J.}~\bibnamefont{Zhang}},
  \bibinfo{author}{\bibfnamefont{K.~S.~M.} \bibnamefont{Reddy}},
  \bibinfo{author}{\bibfnamefont{S.~D.} \bibnamefont{Flexner}},
  \bibinfo{author}{\bibfnamefont{C.~J.} \bibnamefont{Palmstrom}},
  \bibnamefont{and} \bibinfo{author}{\bibfnamefont{P.~A.}
  \bibnamefont{Crowell}}, \bibinfo{journal}{Nat Phys}
  \textbf{\bibinfo{volume}{3}}, \bibinfo{pages}{197} (\bibinfo{year}{2007}).

\bibitem[{\citenamefont{Frolov et~al.}(2009)\citenamefont{Frolov, Venkatesan,
  Yu et~al.}}]{Frolov09}
\bibinfo{author}{\bibfnamefont{S.~M.} \bibnamefont{Frolov}},
  \bibinfo{author}{\bibfnamefont{A.}~\bibnamefont{Venkatesan}},
  \bibinfo{author}{\bibfnamefont{W.}~\bibnamefont{Yu}},
  \bibinfo{author}{\bibfnamefont{J.~A.} \bibnamefont{Folk}}, \bibnamefont{and}
  \bibinfo{author}{\bibfnamefont{W.}~\bibnamefont{Wegscheider}},
  \bibinfo{journal}{Phys. Rev. Lett.} \textbf{\bibinfo{volume}{102}},
  \bibinfo{pages}{116802} (\bibinfo{year}{2009}).

\bibitem[{\citenamefont{Koralek et~al.}(2009)\citenamefont{Koralek, Weber,
  Orenstein et~al.}}]{Koralek09}
\bibinfo{author}{\bibfnamefont{J.~D.} \bibnamefont{Koralek}},
  \bibinfo{author}{\bibfnamefont{C.~P.} \bibnamefont{Weber}},
  \bibinfo{author}{\bibfnamefont{J.}~\bibnamefont{Orenstein}},
  \bibinfo{author}{\bibfnamefont{B.~A.} \bibnamefont{Bernevig}},
  \bibinfo{author}{\bibfnamefont{S.-C.} \bibnamefont{Zhang}},
  \bibinfo{author}{\bibfnamefont{S.}~\bibnamefont{Mack}}, \bibnamefont{and}
  \bibinfo{author}{\bibfnamefont{D.~D.} \bibnamefont{Awschalom}},
  \bibinfo{journal}{Nature} \textbf{\bibinfo{volume}{458}},
  \bibinfo{pages}{610} (\bibinfo{year}{2009}).

\bibitem[{\citenamefont{Lechner et~al.}(2009)\citenamefont{Lechner, Golub,
  Olbrich et~al.}}]{Lechner09}
\bibinfo{author}{\bibfnamefont{V.}~\bibnamefont{Lechner}},
  \bibinfo{author}{\bibfnamefont{L.~E.} \bibnamefont{Golub}},
  \bibinfo{author}{\bibfnamefont{P.}~\bibnamefont{Olbrich}},
  \bibinfo{author}{\bibfnamefont{S.}~\bibnamefont{Stachel}},
  \bibinfo{author}{\bibfnamefont{D.}~\bibnamefont{Schuh}},
  \bibinfo{author}{\bibfnamefont{W.}~\bibnamefont{Wegscheider}},
  \bibinfo{author}{\bibfnamefont{V.~V.} \bibnamefont{Bel'kov}},
  \bibnamefont{and} \bibinfo{author}{\bibfnamefont{S.~D.}
  \bibnamefont{Ganichev}} (\bibinfo{year}{2009}), \eprint{arXiv:0903.1232}.

\bibitem[{\citenamefont{Studer et~al.}(2009)\citenamefont{Studer, Salis,
  Ensslin et~al.}}]{Studer09}
\bibinfo{author}{\bibfnamefont{M.}~\bibnamefont{Studer}},
  \bibinfo{author}{\bibfnamefont{G.}~\bibnamefont{Salis}},
  \bibinfo{author}{\bibfnamefont{K.}~\bibnamefont{Ensslin}},
  \bibinfo{author}{\bibfnamefont{D.~C.} \bibnamefont{Driscoll}},
  \bibnamefont{and} \bibinfo{author}{\bibfnamefont{A.~C.}
  \bibnamefont{Gossard}} (\bibinfo{year}{2009}), \eprint{arXiv:0903.0920}.

\bibitem[{\citenamefont{Schliemann et~al.}(2003)\citenamefont{Schliemann,
  Egues, and Loss}}]{Schliemann03}
\bibinfo{author}{\bibfnamefont{J.}~\bibnamefont{Schliemann}},
  \bibinfo{author}{\bibfnamefont{J.~C.} \bibnamefont{Egues}}, \bibnamefont{and}
  \bibinfo{author}{\bibfnamefont{D.}~\bibnamefont{Loss}},
  \bibinfo{journal}{Phys. Rev. Lett.} \textbf{\bibinfo{volume}{90}},
  \bibinfo{eid}{146801} (\bibinfo{year}{2003}).

\bibitem[{\citenamefont{Bychkov and Rashba}(1984)}]{Bychkov84}
\bibinfo{author}{\bibfnamefont{Y.~A.} \bibnamefont{Bychkov}} \bibnamefont{and}
  \bibinfo{author}{\bibfnamefont{E.~I.} \bibnamefont{Rashba}},
  \bibinfo{journal}{J. Phys. C} \textbf{\bibinfo{volume}{17}},
  \bibinfo{pages}{6039} (\bibinfo{year}{1984}).

\bibitem[{\citenamefont{Dresselhaus}(1955)}]{Dresselhaus55}
\bibinfo{author}{\bibfnamefont{G.}~\bibnamefont{Dresselhaus}},
  \bibinfo{journal}{Phys. Rev.} \textbf{\bibinfo{volume}{100}},
  \bibinfo{pages}{580} (\bibinfo{year}{1955}).

\bibitem[{\citenamefont{Winkler}(2003)}]{Winkler03}
\bibinfo{author}{\bibfnamefont{R.}~\bibnamefont{Winkler}},
  \emph{\bibinfo{title}{Spin-Orbit Coupling Effects in Two-Dimensional Electon
  and Hole Systems}} (\bibinfo{publisher}{Springer}, \bibinfo{address}{Berlin},
  \bibinfo{year}{2003}).

\bibitem[{\citenamefont{Shafir et~al.}(2004)\citenamefont{Shafir, Shen, and
  Saikin}}]{shafir04}
\bibinfo{author}{\bibfnamefont{E.}~\bibnamefont{Shafir}},
  \bibinfo{author}{\bibfnamefont{M.}~\bibnamefont{Shen}}, \bibnamefont{and}
  \bibinfo{author}{\bibfnamefont{S.}~\bibnamefont{Saikin}},
  \bibinfo{journal}{Phys. Rev. B} \textbf{\bibinfo{volume}{70}},
  \bibinfo{eid}{241302(R)} (\bibinfo{year}{2004}).

\bibitem[{\citenamefont{Adagideli et~al.}(2009)\citenamefont{Adagideli,
  Bardarson, and Jacquod}}]{Adagideli09}
\bibinfo{author}{\bibfnamefont{I.}~\bibnamefont{Adagideli}},
  \bibinfo{author}{\bibfnamefont{J.~H.} \bibnamefont{Bardarson}},
  \bibnamefont{and} \bibinfo{author}{\bibfnamefont{P.}~\bibnamefont{Jacquod}},
  \bibinfo{journal}{J Phys : Condens Matter} \textbf{\bibinfo{volume}{21}},
  \bibinfo{pages}{155503} (\bibinfo{year}{2009}).

\bibitem[{\citenamefont{Mehta}(2004)}]{Mehta04}
\bibinfo{author}{\bibfnamefont{M.~L.} \bibnamefont{Mehta}},
  \emph{\bibinfo{title}{Random Matrices}} (\bibinfo{publisher}{Elsevier},
  \bibinfo{year}{2004}).

\bibitem[{\citenamefont{Veillette et~al.}(2004)\citenamefont{Veillette, Bena,
  and Balents}}]{Veillette04}
\bibinfo{author}{\bibfnamefont{M.~Y.} \bibnamefont{Veillette}},
  \bibinfo{author}{\bibfnamefont{C.}~\bibnamefont{Bena}}, \bibnamefont{and}
  \bibinfo{author}{\bibfnamefont{L.}~\bibnamefont{Balents}},
  \bibinfo{journal}{Phys. Rev. B} \textbf{\bibinfo{volume}{69}},
  \bibinfo{pages}{075319} (\bibinfo{year}{2004}).

\bibitem[{\citenamefont{Tang et~al.}(2002)\citenamefont{Tang, Monzon, Jedema
  et~al.}}]{Tang02}
\bibinfo{author}{\bibfnamefont{H.~X.} \bibnamefont{Tang}},
  \bibinfo{author}{\bibfnamefont{F.~G.} \bibnamefont{Monzon}},
  \bibinfo{author}{\bibfnamefont{F.~J.} \bibnamefont{Jedema}},
  \bibinfo{author}{\bibfnamefont{A.~T.} \bibnamefont{Filip}},
  \bibinfo{author}{\bibfnamefont{B.~J.} \bibnamefont{van Wees}},
  \bibnamefont{and} \bibinfo{author}{\bibfnamefont{M.~L.}
  \bibnamefont{Roukes}}, \emph{\bibinfo{title}{Semiconductor Spintronics and
  Quantum Computation}} (\bibinfo{publisher}{Springer-Verlag},
  \bibinfo{year}{2002}), chap.~\bibinfo{chapter}{2}, p.~\bibinfo{pages}{68}.

\bibitem[{\citenamefont{Sun et~al.}(2008)\citenamefont{Sun, Xie, and
  Wang}}]{Sun08}
\bibinfo{author}{\bibfnamefont{Q.~F.} \bibnamefont{Sun}},
  \bibinfo{author}{\bibfnamefont{X.~C.} \bibnamefont{Xie}}, \bibnamefont{and}
  \bibinfo{author}{\bibfnamefont{J.}~\bibnamefont{Wang}},
  \bibinfo{journal}{Phys. Rev. B} \textbf{\bibinfo{volume}{77}},
  \bibinfo{eid}{035327} (\bibinfo{year}{2008}).

\bibitem[{\citenamefont{Malajovich et~al.}(2000)\citenamefont{Malajovich,
  Kikkawa, Awschalom et~al.}}]{Malajovich00}
\bibinfo{author}{\bibfnamefont{I.}~\bibnamefont{Malajovich}},
  \bibinfo{author}{\bibfnamefont{J.~M.} \bibnamefont{Kikkawa}},
  \bibinfo{author}{\bibfnamefont{D.~D.} \bibnamefont{Awschalom}},
  \bibinfo{author}{\bibfnamefont{J.~J.} \bibnamefont{Berry}}, \bibnamefont{and}
  \bibinfo{author}{\bibfnamefont{N.}~\bibnamefont{Samarth}},
  \bibinfo{journal}{Phys. Rev. Lett.} \textbf{\bibinfo{volume}{84}},
  \bibinfo{pages}{1015} (\bibinfo{year}{2000}).

\bibitem[{\citenamefont{Datta}(1995)}]{datta}
\bibinfo{author}{\bibfnamefont{S.}~\bibnamefont{Datta}},
  \emph{\bibinfo{title}{Electronic Transport in Mesoscopic Systems}}
  (\bibinfo{publisher}{Cambridge University Press}, \bibinfo{year}{1995}).

\bibitem[{\citenamefont{Rashba}(2000)}]{Rashba00}
\bibinfo{author}{\bibfnamefont{E.~I.} \bibnamefont{Rashba}},
  \bibinfo{journal}{Phys. Rev. B} \textbf{\bibinfo{volume}{62}},
  \bibinfo{pages}{R16267} (\bibinfo{year}{2000}).

\bibitem[{\citenamefont{Beenakker}(1997)}]{Beenakker97}
\bibinfo{author}{\bibfnamefont{C.~W.~J.} \bibnamefont{Beenakker}},
  \bibinfo{journal}{Rev. Mod. Phys.} \textbf{\bibinfo{volume}{69}},
  \bibinfo{pages}{731} (\bibinfo{year}{1997}).

\bibitem[{\citenamefont{B\"uttiker}(1986)}]{Buttiker86}
\bibinfo{author}{\bibfnamefont{M.}~\bibnamefont{B\"uttiker}},
  \bibinfo{journal}{Phys. Rev. Lett.} \textbf{\bibinfo{volume}{57}},
  \bibinfo{pages}{1761} (\bibinfo{year}{1986}).

\bibitem[{\citenamefont{Zumb\"uhl et~al.}(2002)\citenamefont{Zumb\"uhl, Miller,
  Marcus et~al.}}]{Zumbuhl02}
\bibinfo{author}{\bibfnamefont{D.~M.} \bibnamefont{Zumb\"uhl}},
  \bibinfo{author}{\bibfnamefont{J.~B.} \bibnamefont{Miller}},
  \bibinfo{author}{\bibfnamefont{C.~M.} \bibnamefont{Marcus}},
  \bibinfo{author}{\bibfnamefont{K.}~\bibnamefont{Campman}}, \bibnamefont{and}
  \bibinfo{author}{\bibfnamefont{A.~C.} \bibnamefont{Gossard}},
  \bibinfo{journal}{Phys. Rev. Lett.} \textbf{\bibinfo{volume}{89}},
  \bibinfo{pages}{276803} (\bibinfo{year}{2002}).

\bibitem[{\citenamefont{Brouwer and Beenakker}(1996)}]{Brouwer96}
\bibinfo{author}{\bibfnamefont{P.~W.} \bibnamefont{Brouwer}} \bibnamefont{and}
  \bibinfo{author}{\bibfnamefont{C.~W.~J.} \bibnamefont{Beenakker}},
  \bibinfo{journal}{J. Math. Phys.} \textbf{\bibinfo{volume}{37}},
  \bibinfo{pages}{4904} (\bibinfo{year}{1996}).

\bibitem[{\citenamefont{Bergmann}(1982)}]{Bergmann82}
\bibinfo{author}{\bibfnamefont{G.}~\bibnamefont{Bergmann}},
  \bibinfo{journal}{Solid State Commun.} \textbf{\bibinfo{volume}{42}},
  \bibinfo{pages}{815} (\bibinfo{year}{1982}).

\bibitem[{\citenamefont{Chakravarty and Schmid}(1986)}]{Chakravarty86}
\bibinfo{author}{\bibfnamefont{S.}~\bibnamefont{Chakravarty}} \bibnamefont{and}
  \bibinfo{author}{\bibfnamefont{A.}~\bibnamefont{Schmid}},
  \bibinfo{journal}{Physics Reports} \textbf{\bibinfo{volume}{140}},
  \bibinfo{pages}{193} (\bibinfo{year}{1986}).

\bibitem[{\citenamefont{Kiselev and Kim}(2005)}]{Kiselev05}
\bibinfo{author}{\bibfnamefont{A.~A.} \bibnamefont{Kiselev}} \bibnamefont{and}
  \bibinfo{author}{\bibfnamefont{K.~W.} \bibnamefont{Kim}},
  \bibinfo{journal}{Phys. Rev. B} \textbf{\bibinfo{volume}{71}},
  \bibinfo{eid}{153315} (\bibinfo{year}{2005}).

\bibitem[{\citenamefont{Zhai and Xu}(2005)}]{Zhai05}
\bibinfo{author}{\bibfnamefont{F.}~\bibnamefont{Zhai}} \bibnamefont{and}
  \bibinfo{author}{\bibfnamefont{H.~Q.} \bibnamefont{Xu}},
  \bibinfo{journal}{Phys. Rev. Lett.} \textbf{\bibinfo{volume}{94}},
  \bibinfo{eid}{246601} (\bibinfo{year}{2005}).

\bibitem[{\citenamefont{Oliver et~al.}(2002)\citenamefont{Oliver, Yamaguchi,
  and Yamamoto}}]{Oliver02}
\bibinfo{author}{\bibfnamefont{W.~D.} \bibnamefont{Oliver}},
  \bibinfo{author}{\bibfnamefont{F.}~\bibnamefont{Yamaguchi}},
  \bibnamefont{and} \bibinfo{author}{\bibfnamefont{Y.}~\bibnamefont{Yamamoto}},
  \bibinfo{journal}{Phys. Rev. Lett.} \textbf{\bibinfo{volume}{88}},
  \bibinfo{pages}{037901} (\bibinfo{year}{2002}).

\bibitem[{\citenamefont{Saraga and Loss}(2003)}]{Saraga03}
\bibinfo{author}{\bibfnamefont{D.~S.} \bibnamefont{Saraga}} \bibnamefont{and}
  \bibinfo{author}{\bibfnamefont{D.}~\bibnamefont{Loss}},
  \bibinfo{journal}{Phys. Rev. Lett.} \textbf{\bibinfo{volume}{90}},
  \bibinfo{pages}{166803} (\bibinfo{year}{2003}).

\bibitem[{\citenamefont{Kolli et~al.}(2009)\citenamefont{Kolli, Benjamin,
  Coello et~al.}}]{Kolli09}
\bibinfo{author}{\bibfnamefont{A.}~\bibnamefont{Kolli}},
  \bibinfo{author}{\bibfnamefont{S.~C.} \bibnamefont{Benjamin}},
  \bibinfo{author}{\bibfnamefont{J.~G.} \bibnamefont{Coello}},
  \bibinfo{author}{\bibfnamefont{S.}~\bibnamefont{Bose}}, \bibnamefont{and}
  \bibinfo{author}{\bibfnamefont{B.~W.} \bibnamefont{Lovett}},
  \bibinfo{journal}{New J. Phys.} \textbf{\bibinfo{volume}{11}},
  \bibinfo{pages}{013018} (\bibinfo{year}{2009}).

\bibitem[{\citenamefont{B\"uttiker}(1988)}]{Buttiker88}
\bibinfo{author}{\bibfnamefont{M.}~\bibnamefont{B\"uttiker}},
  \bibinfo{journal}{IBM J. Res. Dev.} \textbf{\bibinfo{volume}{32}},
  \bibinfo{pages}{63} (\bibinfo{year}{1988}).

\bibitem[{\citenamefont{Brouwer and Beenakker}(1995)}]{Brouwer95}
\bibinfo{author}{\bibfnamefont{P.~W.} \bibnamefont{Brouwer}} \bibnamefont{and}
  \bibinfo{author}{\bibfnamefont{C.~W.~J.} \bibnamefont{Beenakker}},
  \bibinfo{journal}{Phys. Rev. B} \textbf{\bibinfo{volume}{51}},
  \bibinfo{pages}{7739} (\bibinfo{year}{1995}).

\bibitem[{\citenamefont{Baranger and Mello}(1995)}]{baranger95}
\bibinfo{author}{\bibfnamefont{H.~U.} \bibnamefont{Baranger}} \bibnamefont{and}
  \bibinfo{author}{\bibfnamefont{P.~A.} \bibnamefont{Mello}},
  \bibinfo{journal}{Phys. Rev. B} \textbf{\bibinfo{volume}{51}},
  \bibinfo{pages}{4703} (\bibinfo{year}{1995}).

\bibitem[{\citenamefont{Huibers et~al.}(1998)\citenamefont{Huibers, Patel,
  Marcus et~al.}}]{Huibers98}
\bibinfo{author}{\bibfnamefont{A.~G.} \bibnamefont{Huibers}},
  \bibinfo{author}{\bibfnamefont{S.~R.} \bibnamefont{Patel}},
  \bibinfo{author}{\bibfnamefont{C.~M.} \bibnamefont{Marcus}},
  \bibinfo{author}{\bibfnamefont{P.~W.} \bibnamefont{Brouwer}},
  \bibinfo{author}{\bibfnamefont{C.~I.} \bibnamefont{Duru\"oz}},
  \bibnamefont{and} \bibinfo{author}{\bibfnamefont{J.~S.}
  \bibnamefont{Harris}}, \bibinfo{journal}{Phys. Rev. Lett.}
  \textbf{\bibinfo{volume}{81}}, \bibinfo{pages}{1917} (\bibinfo{year}{1998}).

\end{thebibliography}

\end{document}